\RequirePackage{fix-cm}
\documentclass[referee]{svjour3}                     
\smartqed  
\usepackage{graphicx}
\usepackage{amssymb}
\usepackage{amsmath}
\usepackage{lineno}
\newcommand{\bg}{\begin{linenomath*}\begin{eqnarray}}
\newcommand{\ed}{\end{eqnarray}\end{linenomath*}}

\newcommand{\lila}{
\text{
$\lim_{\dl\downarrow 0}$
}
}

\newcommand{\dl}{
\text{
\hspace{-0.2cm}
$\lambda$
\hspace{-0.2cm}
}
}
\newcommand{\sidl}{
\text{
$\sin(\dl)$
}
}
\newcommand{\codl}{
\text{
$\cos(\dl)$
}
}
\newcommand{\snx}{
\text{
$
\sin
\left(
\frac{x}{2}
\right)
$
}
}
\newcommand{\cox}{
\text{
$
\cos
\left(
\frac{x}{2}
\right)
$
}
}
\newcommand{\apr}{
\text{
\hspace{-0.2cm}
$\mathcal{A}_p$
}
}

\newcommand{\out}{
\text{
$\hat{n}$
}
}
\newcommand{\an}
{
\text{an}
}
\newcommand{\noan}
{
\text{ok}
}
\newcommand{\atg}
{
\text{
arctan
}
}
\begin{document}

\title{Wigner's effective mathematics and contradiction 
}


\author{
Han Geurdes
}
\institute{GDS Applied Math BV 2593 NN 164, Den Haag Netherlands, han.geurdes@gmail.com}

\date{Received: date / Accepted: date}

\maketitle

\begin{abstract}
Complex numbers are basic.
An inconsistency would question Wigner's unreasonable effectiveness of mathematics.
A vehicle to study this question is Kirchoff's scalar diffraction theory.
In the paper, an inconsistency in complex phase angle is presented. 
When this inconsistency is introduced in Kirchoff's theory we can study its influence on the experimental success of this theory.
There are no \emph{a priori} reasons to include or exclude phase angles. 
Referring to Wigner, an experiment can provide more insight.
In the experiment a weak intensity, small wavelength source can be employed. When the contradictory phase angle is excluded, a nonzero diffraction amplitude appears physically possible. 
If it is included, this amplitude vanishes.
\end{abstract}
\acknowledgement{The author wishes to acknowledge the support of Ad Popper, director Xilion BV.}
\section{Introduction}
The contradiction in complex numbers is presented first.
A paper with details  viz. a preprint \cite{Han}, is already under review. 
Secondly, the contradiction is introduced in Kirchhoff's diffraction.
Kirchhoff's theory is inconsistent \cite[pg. 45-46]{Goodman} and isn't a first approximation either \cite{MW}, \cite{Saatsi}. 
It is however successful in describing diffraction  \cite[pg 482]{Jack}.
Finally the mathematical predictions of the experiment are presented.
Wigner's unreasonable effectiveness of mathematics  \cite{Wign} is the theme behind the exercise.
There is no existing literature for the present study.
Only a few excellent textbooks and old papers were consulted. 

\section{Contradiction \& Kirchhoff diffraction}
Let us start with 
\bg\label{An1}
f_{\dl}(y)
=
e^{iy}
\left\{
e^{i\dl} -(1+\sin(\dl))
\right\}
\ed
under $\dl\rightarrow 0^{+}$.
Here $y=\left(\frac{x+\pi}{2}\right) \in \mathbb{R}$, $x\in\mathbb{R}$.
Equation (\ref{An1}) is written $f_{\dl}(y)=|f_{\dl}(y)|\exp(i\varphi_{\dl})$.
With $|f_{\dl}(y)|^2=1+(1+\sin(\dl))^2-2\cos(\dl)(1+\sin(\dl))$.
Applying a number of times the rule of lHopital, the reader can check that
\bg\label{An2}
\lila \frac{\sin^2(\dl)}{|f_{\dl}(y)|^2}=\frac{1}{2}
\ed
viz. \cite{Han}.
Similarly, noting $\sin(\lambda)\propto \lambda $ for small $\lambda$,
\bg\label{An2x}
\lila \frac{|f_{\dl}(y)|^2}{\lambda^2} =2
\ed
Further, with Euler's identity \cite{Desh}, $e^{i\theta}=\cos(\theta)+i\sin(\theta)$, $\theta \in\mathbb{R}$, equation (\ref{An1}) is written 
\bg\label{An3}
\cos(y+\dl)-(1+\sin(\dl))\cos(y) = |f_{\dl}(y)|\cos(\varphi_{\dl})\\\nonumber
\sin(y+\dl)-(1+\sin(\dl))\sin(y) = |f_{\dl}(y)|\sin(\varphi_{\dl})
\ed
Then the first equation equation (\ref{An3}) is
\bg\label{An5}
\cos(\varphi_{\dl})
=
-
\snx
\frac{\codl-1}{\sin(\dl)}
\left(
\frac{\sin(\dl)}{|f_{\dl}(y)|}
\right)
+\frac{\sidl}{|f_{\dl}(y)|}
\left(
\snx-\cox
\right)
\ed
Now, with $\cos
\left(
\frac{x+\pi}{2}
\right)=-\sin(\frac{x}{2})$, the first term on the right hand side of equation (\ref{An5}) vanishes because,
\bg\label{An6}
\lila \frac{\codl-1}{\sin(\dl)} =0
\ed
From equation (\ref{An2}) and the existence of $\varphi=\lila \varphi_{\dl}$ it follows 
\bg\label{An7}
\cos(\varphi)=
\frac{1}{\sqrt{2}}
\left(
\snx-\cox
\right)
\ed
This is our first equation of concern.
The second equation of (\ref{An3})
is
\bg\label{An8}
\sin(\varphi_{\dl})
=
\sin
\left(
\frac{x+\pi}{2}
\right)
\frac{\cos(\dl)}{|f_{\dl}(y)|}
+
\cos
\left(
\frac{x+\pi}{2}
\right) 
\frac{\sin(\dl)}{|f_{\dl}(y)|}
-\frac{(1+\sin(\dl))}
{|f_{\dl}(y)|}
\sin
\left(
\frac{x+\pi}{2}
\right)
\ed
And so, with equation (\ref{An6}), equation (\ref{An2}), $\sin
\left(
\frac{x+\pi}{2}
\right)=\cos(\frac{x}{2})$ and along a similar path as previously, we derive the second equation of concern
\bg\label{An9}
\sin(\varphi)=
-\frac{1}{\sqrt{2}}
\left(
\snx+\cox
\right) 
\ed
In Table-\ref{Tab1} for $0\leq x \leq 3\pi/2$, a number of points $\varphi_{\dl}$ was computed.
The result presented in Table-1 below extends what has been written in the preprints \cite{Han}. 
For example, $x=2\pi/3$, then equations (\ref{An5}) and (\ref{An9}) give
$
\cos(\varphi)
\approx 0.259
$
and
$
\sin(\varphi)
\approx -0.966
$
and from equations (\ref{An5}) and (\ref{An9}) $
\sin(2\varphi)=2\cos(\varphi)\sin(\varphi)=- \frac{1}{2}
$
together with
$
\cos(2\varphi)=\cos^2(\varphi)-\sin^2(\varphi)=-\frac{\sqrt{3}}{2}
$.
With $\varphi^{\an}=7\pi/12$, which is an allowed choice, we then find
$
\cos(\varphi)=
\cos
\left(
\frac{7\pi}{12}
\right)
\approx -0.259
$
and
$
\sin(\varphi)=
\sin
\left(
\frac{7\pi}{12}
\right)
\approx 0.966
$.
\newline
\hspace*{5cm} INSERT Table-1 about here.
\newline
The counter argumentation where one starts with $q=-1$ and then computes $q^2=1$ and subsequently has two different solutions is \emph{not} equivalent to what is presented here. 
The difference is that with $q=-1$, a correct soltion is primed first.
In our analysis no correct solution is primed first.

\section{Scalar diffraction}
In the \emph{second} place let us look at Kirchhoff's scalar diffraction \cite{Goodman} with small $\lambda$.
The geometry of the screen plus aperture is standard and can be found in e.g. \cite[pg 45, fig 3.7]{Goodman}.
X-ray diffraction is e.g. applied in physical chemistry \cite{Kovar}. Kirchhoff's diffraction theory is also applied in acoustic holography  \cite{Vibr}.

In scalar diffraction theory, the field vector entries in $P_0=(x_{01},x_{02},x_{03})=\vec{x}_0$ at time $t$ are written as $u(P_0 , t)=\Re_e\{U(P_0)\exp[-2i\pi t \nu]\}$, viz. \cite[pg 38, eq 3-10]{Goodman}. 
The expression for $U(P_0)$ is then in Kirchoff's theory a surface integral over the aperture $\apr$ 
\bg\label{K1}
U(P_0)=\frac{1}{4\pi} \iint_{\apr} 
\left\{
G(P_1)\frac{\partial}{\partial n}U(P_1) -
U(P_1)\frac{\partial}{\partial n}G(P_1)
\right\}dS
\ed
The $U(P_1)$ is a spherical wave that from $P_2$, illuminates the screen plus aperture \cite[pg 45, fig 3.7]{Goodman}.
Note that $\frac{\partial}{\partial n} = \hat{n}.\nabla_1$. $\out$ is the outward directed (towards $P_2$) normal, $||\out||=1$, of the aperture $\apr$ and $\nabla_1=\left(
\frac{\partial}{\partial x_{11}},
\frac{\partial}{\partial x_{12}},
\frac{\partial}{\partial x_{13}}
\right)$.
The $||\cdot||$ is the euclidean norm.
The  function $U(P_1)$ solves the Helmholtz equation $\left(\nabla_1^2 +k^2\right)U(P_1)=0$, with $k$ the wave number $k=2\pi / \lambda$.
\bg\label{K2}
U(P_1)=A_{\dl}\frac{\exp[ikr_{21}]}{r_{21}}
\ed
Here, $A=A_{\dl}=\left(
e^{i\lambda}-(1+\sin(\lambda))
\right)$ is a constant in $\vec{x}_1$ and $\vec{x}_2$, viz. \cite[pg 45]{Goodman}.
The $r_{21}$ is the Euclidean distance between point $P_2$ and $P_1$ in the aperture, or $r_{21}=||\vec{x}_1-\vec{x}_2||>0$.
Let us subsequently define the Green function in equation (\ref{K1}) as in Kirchhoff's theory \cite[pg43]{Goodman}
\bg\label{K3}
G(P_1)=
\frac{\exp[ikr_{01}]}{r_{01}}
\ed
$P_2$ and $P_0$ are at opposite sides of the screen. 
They are not necessarily "mirror" images.
The Green function, $G(P_1)$, also solves the Helmholtz equation (in $\vec{x}_1$).
Furthermore,  $r_{01}=||\vec{x}_1-\vec{x}_0||$ and  
\bg\label{K3a}
g_{\,\dl}(P_1)=\exp[ikr_{01}]\left(
e^{i\lambda}-(1+\sin(\lambda))
\right)
\ed
The $y$ in equation (\ref{An1}) is $kr_{01}$.
Hence, $y=(x+\pi)/2$  equation (\ref{K3a}), is related to $kr_{01}$. 
When, $\ell'\in\mathbb{N}$ then, $y-2\pi\ell'$ is equivalent to $y$ in the analysis. 
From $0\leq x \leq 3\pi/2$ as in Table-\ref{Tab1}, it is possible to obtain: $\dl(\ell'+\frac{1}{4})\leq r_{01}\leq \dl(\ell' + \frac{5}{8})$. 
In this way the set of possible $r_{01}$ embraces realistic observer positions $P_0$.
For example, in the visible range e.g. $\dl=4\times 10^{-7}$ meter. If, $\ell'=10^{7}$ then: $4+10^{-7}\leq r_{01}\leq 4+(\frac{20}{8}\times 10^{-7})$ determines $r_{01}$ within Table-\ref{Tab1}.
Table-\ref{Tab1} can be extended.

Subsequently,
\bg\label{K4}
\frac{\partial}{\partial n}
U(P_1)
=\cos(\out,\vec{x}_{21})
\left(
ik-\frac{1}{r_{21}}
\right)
U(P_1)
=
\\\nonumber
\cos(\out,\vec{x}_{21})
\sqrt{
\left(
\frac{1}{r_{21}^2}+k^2
\right)
}
\,
\exp\left[
-i \atg(kr_{21})
\right]
U(P_1)
\ed
From the inner product of $\out$ and $\vec{x}_{21}$ the cosine $\cos(\out,\vec{x}_{21})=(\out\cdot\vec{x}_{21})/r_{21}$ follows.
The $\cos(\out,\vec{x}_{21})$ is a shorthand for $\cos[\measuredangle(\out,\vec{x}_{21})]$ and $\measuredangle(\out,\vec{x}_{21})$ the angle between $\out$ and $\vec{x}_{21}$. 
Similar to equation (\ref{K4})
\bg\label{K5}
\frac{\partial}{\partial n}
G(P_1)
=
\cos(\out,\vec{x}_{01})
\sqrt{
\left(
\frac{1}{r_{01}^2}+k^2
\right)
}
\,
\exp\left[
-i \atg(kr_{01})
\right]
G(P_1)
\ed
When $\dl \approx 0^{+}$ it follows that 
$\sqrt{
\left(
\frac{1}{r^2}+k^2
\right)
}
\approx k
$
for both $r=r_{01}$ as well as for $r=r_{21}$, with $1/r$ such that $k^2\gg \frac{1}{r^2}$.
Then looking at equations equation (\ref{K3}) and equation (\ref{An2x}) under $\dl \approx 0^{+}$ while  $U(P_1)\frac{\partial}{\partial n} G(P_1)$ as well as in $G(P_1)\frac{\partial}{\partial n} U(P_1)$ contains, referring to equation (\ref{K3a}), the term $k|g_{\,\dl}(P_1)|$.
Therefore
\bg\label{K6}
k|g_{\,\dl}(P_1)|\approx 2\pi\sqrt{2}
\ed
in equation (\ref{K1}).
If we subsequently have nonzero finite $r_{21}$ and $r_{01}$ then $\atg(kr_{21})\approx \frac{\pi}{2}$, and $\atg(kr_{01})\approx \frac{\pi}{2}$ when $\dl\approx 0^{+}$. 
Therefore 
\bg\label{K7}
U(P_0)\approx \frac{\sqrt{2}}{2}
\iint_{\apr}  
\frac{\cos(\out,\vec{x}_{21})-\cos(\out,\vec{x}_{01})}{r_{01}r_{21}}
\exp
\left[
i
\left(
\varphi -\frac{\pi}{2}+kr_{21}
\right)
\right]dS
\ed
With $u(P_0 , t)=\Re_e\{U(P_0)\exp[-2i\pi t \nu]\}$ and $2\pi t \nu = k\frac{tc}{n}$ it follows
\bg\nonumber
u(P_0,t)
\approx
\frac{\sqrt{2}}{2}
\Re_e
\left\{
\iint_{\apr} 
\frac{\cos(\out,\vec{x}_{21})-\cos(\out,\vec{x}_{01})}{r_{01}r_{21}}
\exp
\left[
i
\left(
\varphi -\frac{\pi}{2}+k\left(r_{21}-\frac{tc}{n}\right)
\right)
\right]dS
\right\}
\\\nonumber
=
\frac{\sqrt{2}}{2}
\iint_{\apr} 
\frac{\cos(\out,\vec{x}_{21})-\cos(\out,\vec{x}_{01})}{r_{01}r_{21}}
\cos
\left[
\varphi -\frac{\pi}{2}+k\left(r_{21}-\frac{tc}{n}\right)
\right]dS
\ed
Now, from the previous section there are \emph{two} different $\varphi$. 
In the second place we may deduce from Table-\ref{Tab1} that $\varphi_{\dl}^{\an}=\varphi_{\dl}^{\noan} + \pi$. 
Then,   
\[
\cos
\left[
\varphi_{\dl}^{\an} -\frac{\pi}{2}+k\left(r_{21}-\frac{tc}{n}\right)
\right]=-\cos
\left[
\varphi_{\dl}^{\noan} -\frac{\pi}{2}+k\left(r_{21}-\frac{tc}{n}\right)
\right].
\]
This implies,
\bg\label{K9}
u(P_0,t)
\approx
\frac{\sqrt{2}}{4}
\iint_{\apr}
\hspace{-0.3cm}\frac{\cos(\out,\vec{x}_{21})-\cos(\out,\vec{x}_{01})}{r_{01}r_{21}}
\cos
\left(
w(\varphi)
\right)
\hspace{0.05cm}
\large{|}_{\varphi=\varphi_{\dl}^{\an} }
\hspace{0.05cm}
dS
\\\nonumber
+
\frac{\sqrt{2}}{4}
\iint_{\apr} 
\hspace{-0.3cm}\frac{\cos(\out,\vec{x}_{21})-\cos(\out,\vec{x}_{01})}{r_{01}r_{21}}
\cos
\left(
w(\varphi)
\right)
\hspace{0.05cm}
\large{|}_{\varphi=\varphi_{\dl}^{\noan} }
\hspace{0.05cm}
dS
\ed
with, $w(\varphi)=\varphi -\frac{\pi}{2}+k\left(r_{21}-\frac{tc}{n}\right)$.
Note that because of  equation (\ref{An2x}), equation (\ref{K9})  is independent of  $A_{\dl}$.
Hence,  when $\varphi_{\dl}^{\an}$ is included,
$
u(P_0,t)
\approx 0
$. 

\section{Result \& discussion}
There is no reason to disallow $\varphi=\varphi_{\dl}^{\an}$ in the first integral of equation (\ref{K9}).
The $\varphi=\varphi_{\dl}^{\an}$ is in all aspects, except for the anomaly, equivalent to $\varphi=\varphi_{\dl}^{\noan}$.
No a priori rule applies and no proper angle is primed first. 
If nature in diffraction
\begin{itemize}
\item{
\emph{includes} $\varphi_{\dl}^{\an}$ then $|u(P_0,t)|\approx 0$, is predicted via equation (\ref{K9}) in experiment.
}
\item{
\emph{excludes} $\varphi_{\dl}^{\an}$ then despite small $A_{\dl}$ in experiment,  
$|u(P_0,t)|>0$
is possible,
}
\end{itemize}
To elaborate the exclusion of $\varphi^{\an}$ somewhat further.
One can add $2\ell\pi$, with $\ell\in\mathbb{N}$, in the $\cos$ argument.
This allows $r_{21}=(tc/n)+\ell\dl$. 
Therefore ($\dl \approx 0^{+}$ and sufficient large $\ell \in \mathbb{N}$)
\bg\label{c1}
u_{excl}(P_0,t) \propto
\frac{\sqrt{2}}{4}
\left(
\frac{1}{(tc/n)+\dl\ell}
\right)
\iint_{\apr}
\left(
\frac{\cos(\out,\vec{x}_{01})-\cos(\out,\vec{x}_{21})}{r_{01}}
\right)
\sin
(
\varphi_{\dl}^{\noan}
)
dS 
\ed
In (\ref{c1}) to ease discussion , the aperture $P_1$ variation doesn't influence the $r_{21}$ much.
At small wavelengths, geometry seems to dominate $|u_{excl}(P_0,t)|$. This agrees with \cite{Keller}.

Furthermore, the fact $r_{21}>0$, entails  $\frac{ct}{n\dl}>\ell$.
Hence, when $\varphi_{\dl}^{\an}$ is \emph{excluded} and $\nu t>\ell$, and given $\Delta \mathcal{E}_m=\hbar \nu_m$, per photon, viz. \cite{Lamb}, provided $\Delta E=\sum_{m=1}^{\ell} \Delta \mathcal{E}_m$ with $\nu_m=\nu$, then, we can have $\Delta E \Delta t > \hbar\ell $. 
In conclusion, a point source $U(P_1)$ in equation (\ref{K2}), will allow $|u_{excl}(P_0,t)|>0$ when coherence $\nu_m=\nu$ occurs for $\ell$ photons and the inconsistent phase angle is excluded in nature.
Looking at equation (\ref{c1}), experimental conditions can most likely be found where $|u_{excl}(P_0,t)|\gg 0$.

Will the contradictory phase angle be excluded in experiment. What does the result tell us about Wigner's unreasonable effectivness of mathematics \cite{Wign}.

\section*{Declarations}
The author has no conflict of interest. 
The author was not funded. 
There is no data associated.

\begin{table}[h!]
\caption{Table representing a sample of the function $\varphi_{\dl}(x)$, with $\dl\downarrow 0$. 
We have, $y_{j1}=\sin(\varphi_{\dl j})$ and  $y_{j2}=\cos(\varphi_{\dl j})$ with $j=1,2$ and $\varphi_{\dl 2}=\varphi_{\dl 1}+\pi$. Only the sinuses of $2\varphi_{\dl 1}$ and $2\varphi_{\dl 2}$ are presented. 
For later purposes: $\varphi_{\dl 1}=\varphi_{\dl}^{\noan}$ \& $\varphi_{\dl 2}=\varphi_{\dl}^{\an}$.}
\begin{tabular}{ccccccccc}
\hline\label{Tab1}
    x  &$\varphi_{\dl 1}$   & $\varphi_{\dl 2}$ & $\sin(2\varphi_{\dl 1})$ &$ \sin(2\varphi_{\dl 2})$ & $y_{11}$ & $y_{12}$ & $y_{21}$ & $y_{22}$ \\ \hline
0.063&	-2.325&	0.817&	0.998&	0.998&	-0.729&	-0.685&	0.729&	0.685\\
1.068&	-1.822&	1.319&	0.482&	0.482&	-0.969&	-0.249&	0.969&	0.249\\
2.094&  -1.309& 1.833&  -0.500& -0.500& -0.966& 0.259&  0.966&  -0.259\\
2.136&	-1.288&	1.854&	-0.536&	-0.536&	-0.960&	0.279&	0.960&	-0.279\\
3.204&	-0.754&	2.388&	-0.998&	-0.998&	-0.685&	0.729&	0.685&	-0.729\\
4.021&	-0.346&	2.796&	-0.637&	-0.637&	-0.339&	0.941&	0.339&	-0.941\\
4.398&	-0.157&	2.985&	-0.309&	-0.309&	-0.156&	0.988&	0.156&	-0.988\\
    \hline
  \end{tabular}
\end{table} 
\end{document}